\documentstyle[12pt]{article}
\newcommand{\beq}{\begin{equation}}
\newcommand{\eeq}{\end{equation}}
\newcommand{\bea}{\begin{eqnarray}}
\newcommand{\eea}{\end{eqnarray}}
\newcommand{\beas}{\begin{eqnarray*}}
\newcommand{\eeas}{\end{eqnarray*}}
\newcommand{\tr}{\mbox{tr}}
\newcommand{\nn}{\nonumber}

\newcommand{\lam}{\lambda}

\newcommand{\dl}{\delta}
\newcommand{\prt}{\partial}
\newcommand{\dg}{\dagger}

\newcommand{\cD}{{\cal D}}

\newcommand{\Gmu}{\Gamma_{\mu}}
\newcommand{\Gnu}{\Gamma_{\nu}}
\newcommand{\gp}{g^{\prime}}
\newcommand{\oh}{\frac{1}{2}}
\newcommand{\limit}{\rightarrow}
\begin{document}
\topmargin 0pt
\oddsidemargin 5mm
\headheight 0pt
\topskip 0mm

\addtolength{\baselineskip}{0.20\baselineskip}
\hfill UT-Komaba 94-4

\hfill March 1994
\begin{center}

\vspace{36pt}
{\large \bf Critical behavior in $c=1$ matrix model \\
  with branching interactions}

\end{center}

\vspace{36pt}

\begin{center}
 Fumihiko Sugino \footnote{Email address: sugino\%tkyvax.decnet@tkyux.phys.
s.u-tokyo.ac.jp}\\
\vspace{3pt}
{\sl Institute of Physics, University of Tokyo, \\
Komaba, Meguro-ku, Tokyo 153, Japan} \\
\vspace{5pt}
   and \\
\vspace{5pt}
 Osamu Tsuchiya \footnote{Email address: otutiya@tansei.cc.u-tokyo.ac.jp } \\
\vspace{3pt}
{\sl Department of Pure and Applied Sciences, University of Tokyo, \\
Komaba, Meguro-ku, Tokyo 153, Japan}

\end{center}

\vspace{36pt}


\begin{center}
{\bf Abstract}
\end{center}

\vspace{12pt}

\noindent

Motivated by understanding the phase structure of $d >1$ strings
we investigate the $c=1$ matrix model with $g' (\tr M(t)^{2})^{2}$
interaction which is the simplest approximation of the model expected
to describe the critical phenomena of the large-$N$ reduced model
of odd-dimensional matrix field theory.
We find three distinct phases: (i) an ordinary $c=1$ gravity phase,
(ii) a branched polymer phase and (iii) an intermediate phase.
Further we can also analyse the one with slightly generalized
$ g^{(2)} (\frac{1}{N}\tr M(t)^{2})^{2} +g^{(3)} (\frac{1}{N}\tr M(t)^{2})^{3}
+ \cdots + g^{(n)} (\frac{1}{N}\tr M(t)^{2})^{n} $
interaction.
As a result the multi-critical versions of the phase (ii) are found.

\vspace{24pt}

\vfill

\newpage

\section{Introduction}

 Matrix models are very powerful tools so far in the analysis
of $d(\leq 1)$-dimensional noncritical strings or two-dimensional
quantum gravity coupled to the conformal matter with
the central charge  $c\leq 1$. They have
given us much information about string susceptibility exponents,
various correlation functions of scaling operators, non-perturbative
structures obtained from double scaling limit and so on \cite{FGZ}.

 But in $d>1$ case the corresponding matrix models cannot be exactly
solvable at the present, thus it is necessary to exploit techniques for
at least approximately solving \cite{DK} \cite{BZ} \cite{HINS}.
Recently Alvarez-Gaum\'{e} et al. tried the
analysis of
higher dimensonal matrix models by using the large-$N$
reduced model \cite{ABC} \cite{DDSW} \cite{GPK1}.
The $d$-dimensional (lattice) matrix field theory we consider is defined by
 in $d=$even case
\bea
Z_{(d=even)} & = & \int \prod_{x} dM(x) \exp \left[-N\sum_{x} \tr
        \left ( \oh\sum_{\mu=1}^{d}(M(x+\mu)-M(x))^{2}
                                 \right. \right.              \nn \\
   &  &~~~~~~~ \left. \left.+\frac{m^{2}}{2}M(x)^{2}+\frac{g}{4}M(x)^{4}
                            \right ) \right],   \label{o1}
\eea
where $x$ is a site on $d$-dimensional hypercubic lattice and $M(x)$ is a
$N\times N$ hermitian matrix. In $d=$odd case, we decompose space-time as
$\mbox{\boldmath $R$}\times (d-1)\mbox{-dim. lattice}$, and the theory is
defined by
\bea
Z_{(d=odd)} & = & \int \prod_{x} \cD M(t,x) \exp \left[-N\sum_{x}\int dt \tr
        \left\{\oh \dot{M}(t,x)^{2}  \right. \right. \nn \\
    &  &~~~~~ + \oh\sum_{\mu=1}^{d-1}(M(t,x+\mu)-M(t,x))^{2}     \nn \\
   &  & ~~~~~\left. \left.+\frac{m^{2}}{2}M(t,x)^{2}
                      +\frac{g}{4}M(t,x)^{4}\right\}  \right],   \label{o2}
\eea
where $t$ is a continuous parameter and $x$ is a site on $(d-1)$-dimensional
lattice. This theory is free from divergences because of an ultra-violet cutoff
(lattice) and an infra-red cutoff (mass $m$). In large-$N$ limit eq.(\ref{o1})
 is reduced to an one-matrix model
\beq
Z_{(d=even)}^{r} = \int dM \exp \left[-N \tr \left(\oh
      \sum_{\mu =1}^{d}(\Gmu M \Gmu^{\dg}-M)^{2}
     +\frac{m^{2}}{2}M^{2} +\frac{g}{4}M^{4}
                                          \right)\right] \label{1-2}
\eeq
and eq.(\ref{o2}) to a matrix quantum mechanics
\bea
Z_{(d=odd)}^{r} & = & \int \cD M(t) \exp \left[-N\int dt \tr
                           \left(\oh\dot{M}(t)^{2} \right.\right.\nn \\
    &   & \left.\left. +\oh\sum_{\mu =1}^{d-1}(\Gmu M(t)\Gmu^{\dg}-M(t))^{2}
    +\frac{m^{2}}{2}M(t)^{2} +\frac{g}{4}M(t)^{4}\right)\right]
                                          \label{1-3}
\eea
where $\Gmu$'s are traceless $SU(N)$ matrices
commuting only up to an element of the center of $SU(N)$,
\beq
 \Gmu\Gnu =Z_{\nu\mu}\Gnu\Gmu, ~~~~~~~Z_{\mu\nu}=e^{2\pi in_{\mu\nu}/N},
\eeq
and the integers $n_{\mu\nu}$ are defined mod $N$ \cite{EN} \cite{D}.
The dimensionality of
the lattice is completely reduced, but as the price of which the twist matrices
$\Gmu$ must be introduced. In spite of this
simplification, due to the twist matrices an angular integration cannot be
performed exactly, so we have
to use some approximation. If we carry out the angle integral term by term
in the expansion of a hopping term
$N\sum_{\mu =1}^{d}\tr(\Gmu M\Gmu^{\dg}M)$ in eq.(\ref{1-2}),
 infinite terms of type $(\sum_{i=1}^{N}\lam_{i}^{k})
(\sum_{i=1}^{N}\lam_{i}^{l})\cdots$ are induced. ($\lam_{i}$'s are eigenvalues
of $M$.) In order to get the knowledge about the theory (\ref{1-2})
Alvarez-Gaum\'{e} et al. investigated the model restricting
the induced terms to finite and suggested the rich phase structure.

  In this paper we are interested in the theory (\ref{1-3}). In order to get
the hints for the phase structure we exactly solve $c=1$ matrix model with
$(\tr M(t)^{2})^{2}$ interaction, which is obtained as a simplest
nontrivial approximation of the result of angular integration in (\ref{1-3}).
The model is described by the action
\beq
S=\int_{-T/2}^{T/2} dt \left[N\tr\left(\oh\dot{M}(t)^{2}+\oh M(t)^{2}
        +gM(t)^{4}\right)+\gp (\tr M(t)^{2})^{2}\right].
\eeq
We can take a continuum limit by remaining $\gp$ fixed and approaching $g$ to
$g_{c}(\gp)$ a point on a critical line in $g-\gp$ plane. Then we find three
distinct phases, which is characterized by the behavior of string
susceptibility
\beq \chi=\lim_{T \limit \infty}\frac{1}{T N^{2}}\frac{\prt ^{2}\ln Z}{\prt
g^{2}}\left|_{\gp:
\mbox{fixed}}\right.,\label{2-1}
\eeq
 (i) a $c=1$ gravity phase ---$\chi\sim 1/\ln (g-g_{c}(\gp))$,\\
 (ii) a branched polymer phase ---$\chi\sim (g-g_{c}(\gp))^{-1/2}$, and \\
(iii) a phase in between (i) and (ii) ---$\chi\sim \ln(g-g_{c})$.
Espesially the phase (iii) is interesting.
It seems that it suggests the existence of a continuum theory of $c>1$ matter
coupled to gravity.

We can also analyze the model with slightly generic interaction
\beq
g^{(2)} (\frac{1}{N}\tr M(t)^{2})^{2} +g^{(3)} (\frac{1}{N}\tr M(t)^{2})^{3}
+ \cdots + g^{(n)} (\frac{1}{N}\tr M(t)^{2})^{n}
\eeq
similarly.

\section{The reduced model for $d=$odd}

  In eq.(\ref{1-3}) the term-by-term angle integral with respect to a hopping
term
$$-N\int dt \tr [\oh \dot{M}(t)^{2}-\sum_{\mu =1}^{d-1}\Gmu M(t)\Gmu^{\dg}M(t)]
$$
induces derivative coupling terms, say
$$\int dt(\sum_{i=1}^{N}\dot{\lam}_{i}(t)^{k})(\sum_{i=1}^{N}\lam_{i}(t)^{l})
               \cdots , $$
as well as the interactions with no derivatives. Since we are interested in
critical (infra-red) properties of the system, we may expect that
the derivative terms are irrelevant. Assuming it, we will consider the $c=1$
matrix model containing no derivative interactions of the form
\beas
S & = & \int dt \left[N\tr\left(\oh \dot{M}(t)^{2}+\oh M(t)^{2}+gM(t)^{4}
                     \right)\right.  \\
  &   & \left. +N^{2}\sum_{k,l,\cdots}g_{k,l,\cdots}\frac{1}{N}\tr M(t)^{k}
            \frac{1}{N}\tr M(t)^{l}\cdots~~~~~~~~\right].
\eeas
   It is very interesting to understand the phase structure of this model.
For the purpose of this paper we shall consider as a simple approximation
the following system
\bea
Z & = & \int \cD M(t) e^{-S} \nn \\
S & = & \int dt \left[N\tr\left(\oh \dot{M}(t)^{2}+\oh M(t)^{2}+gM(t)^{4}
                     \right) +\gp (\tr M(t)^{2})^{2}\right]  \label{1-4}
\eea
and solve it exactly in $N \limit \infty$ limit.

 We can give a geometrical interpretation for the term
$\gp (\tr M(t)^{2})^{2}$,
similar to the $c=0$ case \cite{DDSW} \cite{GPK1}.
It provides a touching (or branching) point
between two surfaces,
where the height $t$ remains unchanged. In general, the term
$$  N^{2}g_{k_{1}\cdots k_{n}}\frac{1}{N}\tr M(t)^{k_{1}}\cdots \frac{1}{N}
             \tr M(t)^{k_{n}}     $$
represents a touching of $n$ surfaces at a common point unchanging the
height. Thus the model (\ref{1-4}) describes a interacting random surface
in one-dimension rather than free surfaces in the case of ordinary matrix
models (containing only a single trace in the action) \cite{K}.

\section{The exact solution of the model in the large-$N$ limit}

Now we obtain the exact solution of the model (\ref{1-4}) in the
large-$N$ limit.
Introducing a collective field
\beq
\phi (x,t) =\frac{1}{N} \tr \dl (x- M(t)),
\label{1-5}
\eeq
and its conjugate momentum $\pi (x,t)$ satisfying the commutator
\beq
[ \phi (x,t), \pi (y,t) ]= i \dl (x-y),
\eeq
the leading order in $N$ of the collective Hamiltonian reads \cite{DJ}
\bea
H & =& {1 \over 2N^{2}} \int dx
\left( \prt_{x} \pi (x) ) \phi (x) (\prt_{x} \pi (x) \right) \nn \\
 & & + N^{2} \int dx \left[ {\pi^{2} \over 6} \phi (x)^{3}
+(\oh x^{2} +gx^{4} ) \phi (x) \right]
+N^{2} \gp \left(\int dx x^{2} \phi (x) \right)^{2} \nn \\
 & & + N^{2} \mu_{F} \left(1- \int dx \phi (x) \right).
\eea
In the last term  $\mu_{F}$ is a lagrange multiplier respect to the constraint
\beq
\int dx \phi (x) =1.
\label{1-6}
\eeq
The saddle point solution for $\phi (x)$ is given by
\beq
\phi_{0} (x) = \frac{1}{\pi} \sqrt{ 2(\mu_{F} - U(x)) },
\label{1-7}
\eeq
where using the second moment of $\phi $
\beq
c = \int dx x^{2} \phi (x)
\label{1-8}
\eeq
the effective potential $U(x)$ is written as
\beq
U(x) = (\oh +2\gp c) x^{2} +gx^{4}.
\label{1-9}
\eeq
The range of $x$ is in the interval $(-x_{-},x_{-})$.
The parameter $x_{-}$ $(x_{+})$ defined as the smaller (bigger) one of
the positive solutions of an equation
\beq
\mu_{F} -U(x) =0
\eeq
with $g<0$.
For a while we should proceed the argument in the case of $g<0$.

Substituting the saddle point value $\phi_{0} (x)$, using the elliptic
integrals $E(k)$,$K(k)$ with the modulus $k$ defined by
\beq
k^{2} = \frac{x_{-}^{2}}{x_{+}^{2}} =
\frac{1+ 4\gp c -\sqrt{(1+4\gp c)^{2} +16 g\mu_{F} } }{ 1+4\gp c +
\sqrt{(1+4\gp c)^{2} +16g \mu_{F} } }
\label{1-10}
\eeq
and introducing the function
\beq
f(k) = \frac{2\sqrt{2}}{3\pi} \left[ K(k) (k^{2} -1) +E(k) (k^{2} +1) \right]
\eeq
we obtain from (\ref{1-6})
\beq
\mu_{F} = \left(\frac{1}{f(k)}\right)^{4/3}
(-g)^{1/3} k^{2}
\label{1-12}
\eeq
and from (\ref{1-8})
\bea
 c &= & \frac{2\sqrt{2}}{15\pi} (-g)^{-1/3}
\left(\frac{1}{f(k)}\right)^{5/3} \nonumber \\
  & & \times \left(K(k)(-k^{4}+3k^{2} -2) +E(k) (2k^{4}-2k^{2}+2) \right).
\label{1-13}
\eea
Of course the formula of the free energy also can be written, but
 the following analysis leads to very tadious calculations.
Fortunately without it we can see the string susceptibility
from eq.(\ref{1-13}) as
below.
The string susceptibility $\chi$ (\ref{2-1}) is nothing but a two point
connected correlator of $\tr M^{4}$ operators
\beq
\chi =\lim_{T\limit \infty } \frac{1}{T}
<\int_{-T/2}^{T/2} dt \tr M(t)^{4}
\int_{-T/2}^{T/2} dt' \tr M(t')^{4} >_{\mbox{conn.}}
\eeq
on the other hand $\partial c/\partial g$ is a connected two point
function of $\tr M^{2}$ and $\tr M^{4}$ operators
\beq
-\frac{\partial c}{\partial g} = \lim_{T \limit \infty}\frac{1}{T}
< \int_{-T/2}^{T/2} dt \tr M(t)^{2}
\int_{-T/2}^{T/2} dt'\tr M(t')^{4}>_{\mbox{conn.}}
\eeq
Thus due to universality, these two quantities must exhibit a same
critical behavior as a connected correlator of two puncture operators.

{}From eqs.(\ref{1-10}) and (\ref{1-13}) we get the relation of
$g$ and $g'$ with $k$
\beq
\frac{2\sqrt{2}}{15\pi} g'  =
 \frac{ (- (-g)^{1/3} f(k)^{5/3}
 + 2(1+k^{2}) (-g) f(k) )}{
 2(k^{4} -k^{2} +1 )E(k)
  + (-k^{4}+3k^{2} -2 )K(k) }.
\label{1-14}
\eeq
This equation determines a curve $g=g(k,g')$ in $g-g'$ plane for any fixed
value of $k$.
In eq.(\ref{1-13}) the singular behavior of $c$ can come from the following
two roots:

(1) the singularity of $K(k)$ when $k\limit 1$

(2) the singularity of the $g$-dependence of $k$ determined by eq.(\ref{1-14})
when
$$\partial g /\partial k |_{g' :\mbox{fixed}}=0.$$

The singularity of type (1) forms a critical line (i) $g=g(1,g')$ in
the phase diagram (Fig.1).
About that of type (2), its critical line (ii) is nothing but an envelope
of a family of the curves with the parameter $k ~~ \{ g=g(k,g')\}_{0<k<1}$.
And the boundary of the critical lines (i) and (ii)
$g=g' = 5\sqrt{5} /36 \sqrt{3}\pi $ forms a critical point (iii) by itself.

We investigate the critical behavior when $k\limit 1$.
As a result of the expansion of $g$ and $k$ about $g_{c} (g') =g(1,g') $
and $1$, eqs.(\ref{1-13}) and (\ref{1-14}) become respectively

\beq
\frac{g-g_{c}}{-g_{c} }  =
\frac{15}{8} \frac{g'-g_{c}}{g'-10g_{c}} k'^{4} \ln \frac{4}{k'}
 - \frac{15}{32} \frac{5g'+g_{c}}{g'-10 g_{c} } k'^{4}
+O\left( k'^{6} \ln \frac{4}{k'} \right),
\eeq

\bea
c &= & \frac{1}{5} (\frac{3\pi}{4\sqrt{2}} )^{2/3} (-g_{c})^{-1/3} \nonumber \\
 & & \times \left[ 1+ \frac{45}{8} \frac{ g_{c}}{g'-10g_{c} }
  k'^{4} \ln \frac{4}{k'}
-\frac{255}{32} \frac{ g_{c} }{g'-10 g_{c}}  k'^{4}
+O \left( k'^{6} \ln \frac{4}{k'} \right) \right]
\eea

where $k'^{2} = 1-k^{2} $.

Threrfore the critical behavior of $\chi$ is
\beq
\chi \sim - \frac{\partial c}{\partial g}
= - \frac{\frac{\partial c}{\partial k'}}{\frac{\partial g}{\partial k'}}
= \left\{
\begin{array}{c} \frac{\mbox{const}}{\ln (g-g_{c}) } ~~ ( g' >g_{c} (g')) \\
 \mbox{const.} \ln (g-g_{c}) ~~ ( g'=g_{c} (g') )
\end{array} \right.  .
\eeq

Here, the case of $g'>g_{c} (g')$ realizes the phase  on the critical line (i)
which is an ordinary $c=1$ gravity phase, and $g'=g_{c} (g')$ case
corresponds to the critical point (iii).
The susceptibility exponent $\gamma_{str}$ defined by
$\chi \sim (g-g_{c})^{-\gamma_{str}} $ is zero in the both phases, however
the logarithmic corrections differently appear.

The behavior near the envelope (ii) is easily seen by expanding $g$ and $k$
 about the values on the envelope $g_{c} (g'), k_{c} (g')$
\beq
g(k,g') -g_{c} = \left. \frac{\partial^{2} g}{\partial (k^{2})^{2}}
\right|_{\begin{array}{c}g';\mbox{fixed} \\
k=k_{c} \end{array} } (k^{2} -k_{c}^{2} )^{2} + \cdots .
\eeq
It can be shown to be
\beq
\chi \sim \frac{\mbox{const}}{\sqrt{g-g_{c} }} ,
\eeq
thus we recognize that $\gamma_{str} =1/2$ and the line (ii) exhibits
branched polymer phase \cite{DKO} \cite{Z}.

The similar analysis can be done for positive $g$ and
the branched polymer phase can be found on the line connected to (ii).

\section{Discussions}

With the view of understanding the phase structures of $d>1$ strings
we have investigated the $c=1$ matrix model with $(\tr M(t)^{2} )^{2}$
interaction and as a result the three phases are found.
Further we can apply the above analysis to the slightly generalized model
defined by the following action
\bea
\lefteqn{S = \int dt\left[N
\tr ( \oh \dot{M} (t)^{2} +\oh M^{2} (t) + g M(t)^{4} ) \right.} \nn \\
 & & \left. +N^{2}\left(g^{(2)} (\frac{1}{N}\tr M(t)^{2})^{2}
     +g^{(3)} (\frac{1}{N}\tr M(t)^{2})^{3}
     + \cdots + g^{(n)} (\frac{1}{N}\tr M(t)^{2})^{n}\right) \right],
\eea
by replacing $2g'c$ in eqs.(\ref{1-9}) and (\ref{1-10}) to
$ 2g^{(2)} c+3 g^{(3)} c^{2} + \cdots +ng^{(n)}c^{n-1} $.
The analogue of eq.(\ref{1-14}) becomes
\beq
 2(-g)^{2/3}\frac{1+k^{2}}{f(k)^{2/3}}=1+2(2g^{(2)}c+
                          \cdots +ng^{(n)}c^{n-1}),     \label{1-38}
\eeq
where the formula of $c$ (\ref{1-13}) holds without any changes.
If the couplings $g^{(1)},\cdots ,g^{(n)} $ are tuned as
\beq
\frac{\partial g}{\partial (k^{2}) } =
\frac{\partial^{2} g}{\partial (k^{2})^{2}} =\cdots
=\frac{\partial^{j} g}{\partial (k^{2})^{j} } =0 \qquad (\mbox{at} ~~ k=k_{c} )
\eeq
and $g$ is analytic with respect to $k^{2}$ near $k=k_{c}$,
then $\partial g/ \partial k^{2} \sim (k^{2} -k_{c}^{2})^{j}
\sim (g-g_{c})^{j/(j+1)} $.
For $k_{c} \neq 1$ (corresponding to the multi-critical version
 of the phase (ii)
in the previous model ) since $c$ is nonsingular, $\chi$ behaves as
\beq
\chi \sim (g-g_{c}) ^{-j/(j+1)} .
\eeq
Here $j$ is an integer which takes $1, \cdots ,n-1$.

Also for the singular behavior as $k\limit 1$, from eq.(\ref{1-13})
\beq
 c=\frac{1}{5}(\frac{3\pi}{4\sqrt{2}})^{2/3}(-g)^{-1/3}\left[1-
     \frac{5}{8}k'^{4}\ln\frac{4}{k'}+\frac{15}{32}k'^{4}+\cdots \right].
                                 \label{1-39}
\eeq
Using this, as a result of the expansion of eq.(\ref{1-38}) around the
critical values we find
\bea
\lefteqn{f_{1}(g^{(2)},\cdots ,g^{(n)})\frac{g-g_{c}}{-g_{c}}
                    +O(k'^{4}\ln\frac{4}{k'} (g-g_{c}))}\nn \\
  & = & -4\tilde{g}_{c}(\frac{1}{8}k'^{4}\ln\frac{4}{k'}+\frac{1}{32}k'^{4}
                    +\cdots) \nn \\
  &   & +10f_{2}(g^{(2)},\cdots,g^{(n)})(-\frac{1}{8}k'^{4}\ln\frac{4}{k'}
              +\frac{3}{32}k'^{4}+\cdots)    \label{1-40}
\eea
where
\bea
f_{1}(g^{(2)},\cdots,g^{(n)}) & = & -4\tilde{g}_{c}
         +\frac{1}{3}\tilde{g}_{c}^{1/3}-\frac{2}{3}\cdot 3\cdot 1
           \tilde{g}^{(3)}\tilde{g}_{c}^{-1/3}-\cdots     \nn \\
    &    &    -\frac{2}{3}n(n-2)\tilde{g}^{(n)}\tilde{g}_{c}^{-(n-2)/3} , \\
f_{2}(g^{(2)},\cdots,g^{(n)}) & = & 2\cdot 1\tilde{g}^{(2)}
                             +3\cdot 2\tilde{g}^{(3)}
                          \tilde{g}_{c}^{-1/3}+\cdots +n(n-1)\tilde{g}^{(n)}
                            \tilde{g}_{c}^{-(n-2)/3} ,
\eea
and for the notational simplicity following symbols are introduced
\beq
 \tilde{g}_{c}=-\frac{3\pi}{4\sqrt{2}}g_{c},~~~~\tilde{g}^{(i)}=
       (\frac{1}{5}\frac{3\pi}{4\sqrt{2}})^{i-1}g^{(i)} ~~~ (i=1,\cdots ,n).
\eeq
{}From eq.(\ref{1-40}) for the generic $g^{(2)},\cdots, g^{(n)}$
$$   g-g_{c}\sim k'^{4}\ln\frac{4}{k'}  $$
which means the $c=1$ gravity phase. If the couplings $g^{(2)},\cdots ,
g^{(n)}$ are tuned as
$$   5f_{2}(g^{(2)},\cdots ,g^{(n)})=-2\tilde{g}_{c} ,  $$
then $g-g_{c}\sim k'^{4}$. Thus one finds the intermediate phase as same as
previous model.

 And the last possibility of the tuning consistent with $g-g_{c}\limit 0$ is
\beq
  f_{1}(g^{(2)},\cdots ,g^{(n)})=0 ~~~\mbox{and}~~
                   5f_{2}(g^{(2)},\cdots ,g^{(n)})=-2\tilde{g}_{c}.\label{1-44}
\eeq
In this case $O(k'^{4}\ln\frac{4}{k'} (g-g_{c}))$ terms in l.h.s. of
eq.(\ref{1-40}) need to be considered and thus it turns out that $g-g_{c}$
behaves as
$$  g-g_{c}\sim \frac{1}{\ln{\frac{4}{k'}}} ,$$
which leads up to additive terms of polynomials of $g-g_{c}$
$$  c\sim \frac{(\mbox{const})}{g-g_{c}}e^{-(\mbox{const})/(g-g_{c})} .$$
Since any order of derivative of $c$ with respect to $g$ is always regular,
it seems that the tuning (\ref{1-44}) can not lead a continuum theory.

  Thus it turns out that the slightly generalized interaction
$ g^{(2)} (\frac{1}{N}tr M(t)^{2})^{2} +g^{(3)} (\frac{1}{N}\tr M(t)^{2})^{3}
+ \cdots + g^{(n)} (\frac{1}{N}\tr M(t)^{2})^{n} $
can make the branched polymer phase multi-critical but can not give any
influences to the $c=1$ gravity phase and the intermediate one.
This situation is similar as that in a following one-matrix model
$$ S=N\tr (\frac{1}{2}M^{2}+gM^{4}) +N^{2}\left\{
 g^{(2)} (\frac{1}{N}\tr M^{2})^{2} + \cdots
 + g^{(n)} (\frac{1}{N}\tr M^{2})^{n} \right\}$$
which is discussed in ref. \cite{ABC}.

   In probing in the possiblities of well-defined $d>1$ continuum string
theory, to investigate deeply the properties of the new phases (ii) and
 (iii) would be very interesting. Related to this point the analysis of the
amplitudes of macroscopic loops is in progress. (In the $d=$even case some
arguments about it are done in ref. \cite{ABC} \cite{GPK2}.)

\vspace{12pt}




\newpage

{\large\bf Figure Captions }

Fig.1:
 Phase diagram of the theory defined by eq.(\ref{1-4}).
The curve is the critical line of the theory.
The piece of the curve (i) belonging to $\gp > 5\sqrt{5} /36\sqrt{3} \pi$
comes from the singularity of type (1) and corresponds to the ordinary $c=1$
gravity phase.
The $\gp < 5\sqrt{5} /36\sqrt{3} \pi $ piece (ii), from the singularity of
the type (2), corresponds to the branched polymer phase.
The point between (i) and (ii) $g=\gp =5\sqrt{5} /36\sqrt{3} \pi $
is a critical point with respect to an intermediate phase.

\end{document}